\documentclass[aps,prd,showpacs,singlecolumn]{revtex4}
\usepackage{amsfonts}
\usepackage{amssymb}
\usepackage{amsmath}
\usepackage{graphicx}
\usepackage{epsfig}

\begin{document}

\title{Complete particle-pair annihilation as a dynamical signature of the
spectral singularity}
\author{G. R. Li, X. Z. Zhang, and Z. Song}
\email{songtc@nankai.edu.cn}
\affiliation{School of Physics, Nankai University, Tianjin 300071, China}

\begin{abstract}
Motivated by the physical relevance of a spectral singularity of interacting
many-particle system, we explore the dynamics of two bosons as well as
fermions in one-dimensional system with imaginary delta interaction
strength. Based on the exact solution, it shows that the two-particle
collision leads to amplitude-reduction of the wave function. For fermion
pair, the amplitude-reduction depends on the spin configuration of two
particles. In both cases, the residual amplitude can vanish when the
relative group velocity of two single-particle Gaussian wave packets with
equal width reaches the magnitude of the interaction strength, exhibiting
complete particle-pair annihilation at the spectral singularity.
\end{abstract}

\pacs{03.65.-w, 03.65.Nk, 11.30.Er}
\maketitle

\section{Introduction}

Non-Hermitian operator has been introduced phenomenologically as an
effective Hamiltonian to fit experimental data in various fields of physics
\cite{Gamow,Dattoli,Okolowicz,Moiseyev,Muga}. In spite of the important role
played non-Hermitian operator in different branches of physics, it has not
been paid due attention by the physics community until the discovery of
non-Hermitian Hamiltonians with parity-time symmetry, which have a real
spectrum \cite{Bender}. It has boosted the research on the complex extension
of quantum mechanics on a fundamental level \cite%
{Ann,JMP1,JPA1,JPA2,PRL1,JMP2,JMP3,JMP4,JPA3,JPA4,JPA5}. Non-Hermitian
Hamiltonian can possess peculiar feature that has no Hermitian counterpart.
A typical one is the spectral singularity (or exceptional point for finite
system), which is a mathematic concept. It has gained a lot of attention
recently \cite{PRA1,PRB1,Ali3,PRA3,JMP5,PRD1,PRA4,PRA5,PRA6}, motivated by
the possible physical relevance of this concept since the pioneer work of\
Mostafazadeh \cite{PRL3}.

The majority of previous works focus on the non-Hermitian system arising
from the complex potential, mean-field nonlinearity \cite%
{PRA2,JPA6,Ali3,PRA13,prd2,prd3,prd4,prd5,prd6,prd7,prd8} as well as
imaginary hopping integral \cite{PRA14}. In this paper, we investigate the
physical relevance of the spectral singularities for non-Hermitian
interacting many-particle system. The non-Hermiticity arises from the
imaginary interaction strength. For two-particle case, the exact solution
shows that there exist a series of spectral singularities, forming a
spectrum of singularity associated with the central momentum of the two
particles. We consider dynamics of two bosons as well as fermions in
one-dimensional system with imaginary delta interaction strength. It shows
that the two-particle collision leads to amplitude-reduction of the wave
function. For fermion pair, the amplitude-reduction depends on the spin
configuration of two particles. Remarkably, in both cases, the residual
amplitude can vanish only when the relative group velocity of two
single-particle Gaussian wave packets with equal width reaches the magnitude
of the interaction strength. This phenomenon of complete particle-pair
annihilation is the direct result of the spectral singularity. We also
discuss the complete annihilations of a singlet fermion pair and a maximally
two-mode entangled boson pair based on the second quantization formalism.

This paper is organized as follows. In Section \ref{Hamiltonian and
solutions}, we present the model Hamiltonian and exact solution. In Section %
\ref{Dynamical signature}, we construct the local boson pair initial state
as initial state which is allowed to calculate the time evolution. Based on
this, we reveal the connection between the phenomenon of complete pair
annihilation and the spectral singularity. In Section \ref{Second
quantization representation}, we extend our study a singlet fermion pair and
a maximally two-mode entangled boson pair based on the second quantization
formalism. Finally, we give a summary in Section \ref{Summary}.

\section{Hamiltonian and solutions}

\label{Hamiltonian and solutions}

We start with an one-dimensional two-distinguishable particle system with
imaginary delta interaction. The solution can be used to construct the
eigenstates of two-fermion and boson systems. The Hamiltonian has the form

\begin{equation}
H_{\mathrm{2p}}=-\frac{1}{2}\left( \frac{\partial ^{2}}{\partial x_{1}^{2}}+%
\frac{\partial ^{2}}{\partial x_{2}^{2}}\right) -i2\gamma \delta \left(
x_{1}-x_{2}\right)
\end{equation}%
where $\gamma >0$\ and we use dimensionless units $e=\hbar =m=1$ for
simplicity.

Introducing new variables $R$ and $r$, where

\begin{eqnarray}
R &=&(x_{1}+x_{2})/2, \\
r &=&x_{1}-x_{2},  \notag
\end{eqnarray}%
we obtain the following Hamiltonian

\begin{equation}
H_{\mathrm{2p}}=H_{\mathrm{R}}+H_{\mathrm{r}},
\end{equation}%
with%
\begin{eqnarray}
H_{\mathrm{R}} &=&-\frac{\partial ^{2}}{4\partial R^{2}}, \\
H_{\mathrm{r}} &=&-\frac{\partial ^{2}}{\partial r^{2}}-i2\gamma \delta
\left( r\right) .  \notag
\end{eqnarray}%
Here $R$\ is the center-of-mass coordinate and $r$\ is the relative
coordinate. The Hamiltonian is separated into a center-of-mass part and a
relative part, and can be solvable exactly.

The eigenfunctions of the center-of-mass motion $H_{\mathrm{R}}$ are simply
plane waves, while the Hamiltonian $H_{\mathrm{r}}$ is equivalent to that of
a single-particle in an imaginary delta-potential, which has been exactly
solved in the Ref.\cite{Ali1}. Then the eigen functions of the original
Hamiltonian can be obtained and expressed as

\begin{eqnarray}
\psi _{+}\left( K,k,x_{1},x_{2}\right) =e^{iK\left( x_{1}+x_{2}\right)
/2}\left\{ \cos \left[ k\left( x_{1}-x_{2}\right) \right] \right. &&
\label{WF_even} \\
\left. -\frac{i\gamma }{k}\sin \left[ k\left( x_{1}-x_{2}\right) \right]
\text{\textrm{sign}}\left( x_{1}-x_{2}\right) \right\} , &&  \notag
\end{eqnarray}%
in symmetrical form, and%
\begin{equation}
\psi _{-}\left( K,k,x_{1},x_{2}\right) =e^{iK\left( x_{1}+x_{2}\right)
/2}\sin \left[ k\left( x_{1}-x_{2}\right) \right] ,  \label{WF_odd}
\end{equation}%
in antisymmetrical form. The corresponding energy is
\begin{equation}
E\left( K,k\right) =K^{2}/4+k^{2},
\end{equation}%
with the central and relative momenta $K,k\in \left( -\infty ,\infty \right)
$. The symmetrical wavefunction $\psi \left( K,k,x_{1},x_{2}\right) $ is the
spatial part wavefunction for two bosons or two fermions in singlet pair,
while the antisymmetrical wavefunction $\varphi \left(
K,k,x_{1},x_{2}\right) $ only for two triplet fermions.

Before starting the investigation on dynamics of two-particle collision, we
would like to point that there exist spectral singularities in the present
Hamiltonian. It arises from the same mechanism as that in the
single-particle systems \cite{Samsonov,Ali1}.

We can see that the eigen functions with even parity and momentum $k=-\gamma
$ can be expressed in the form%
\begin{eqnarray}
\psi _{\text{ss}}\left( K\right) &\equiv &\psi \left( K,-\gamma
,x_{1},x_{2}\right) \\
&=&e^{iK\left( x_{1}+x_{2}\right) /2}e^{-i\gamma \left\vert
x_{1}-x_{2}\right\vert },  \notag
\end{eqnarray}%
with energy
\begin{equation}
E_{\text{ss}}\left( K\right) =K^{2}/4+\gamma ^{2}.
\end{equation}%
We note that function $\psi _{\text{ss}}\left( K\right) $\ satisfies%
\begin{equation}
\underset{x_{1}-x_{2}\rightarrow \pm \infty }{\lim }\left[ \frac{\partial
\psi _{\text{ss}}\left( K\right) }{\partial \left( x_{1}-x_{2}\right) }\pm
i\gamma \psi _{\text{ss}}\left( K\right) \right] =0,
\end{equation}%
which accords with the definition of the spectral singularity in Ref. \cite%
{Ali3}. It shows that there exist a series of spectral singularities
associated with energy $E_{\text{ss}}\left( K\right) $ for $K\in \left(
-\infty ,\infty \right) $, which constructs a spectrum of spectral
singularities. We will demonstrate in the following section that such a
singularity spectrum leads to a peculiar dynamical behavior of two local
boson pair or equivalently, singlet fermion pair.

\section{Dynamical signature}

\label{Dynamical signature}

\subsection{Construction of initial state}

The emergence of the spectral singularity induces a mathematical obstruction
for the calculation of the time evolution of a given initial state, since it
spoils the completeness of the eigen functions and prevents the eigenmode
expansion. Nevertheless, the completeness of the eigen functions is not
necessary for the time evolution of a state with a set of given coefficients
of expansion. It does not cause any difficulty in deriving the time
evolution of an initial state with arbitrary combination of the eigen
functions. Namely, any linear combination of function set $\left\{ \psi
\left( K,k,x_{1},x_{2}\right) \right\} $\ or $\left\{ \varphi \left(
K,k,x_{1},x_{2}\right) \right\} $\ can be an initial state, and the time
evolution of it can be obtained simply by adding the factor $e^{-iE\left(
K,k\right) t}$.

In order to investigate the dynamical consequence of the singularity
spectrum, we consider the time evolution of the initial state of the form

\begin{equation}
\Psi \left( x_{1},x_{2},0\right) =\frac{1}{\sqrt{\Lambda }}\int_{-\infty
}^{\infty }\int_{-\infty }^{\infty }G\left( K\right) g\left( k\right) \psi
\left( K,k,x_{1},x_{2}\right) \text{\textrm{d}}K\text{\textrm{d}}k,
\end{equation}%
where $\Lambda $ is the normalization factor, which will be given in the
following and

\begin{eqnarray}
G\left( K\right) &=&\exp \left[ -\frac{1}{2\alpha ^{2}}(K-K_{0})^{2}-iKR_{0}%
\right] , \\
g\left( k\right) &=&\exp \left[ -\frac{1}{2\beta ^{2}}(k-k_{0})^{2}-ikr_{0}%
\right] .
\end{eqnarray}%
Here $\alpha ,\beta ,r_{0},k_{0}>0$ and $K_{0}$ is arbitrary real number. We
explicitly have%
\begin{equation}
\Psi \left( x_{1},x_{2},0\right) =\frac{\pi \alpha \beta }{2\sqrt{\Lambda }%
k_{0}}\left[ \left( k_{0}+\gamma \right) \exp \theta _{+}+\left(
k_{0}-\gamma \right) \exp \theta _{-}\right]
\end{equation}%
where%
\begin{eqnarray}
\theta _{\pm } &=&-\frac{\alpha ^{2}}{2}\left( \frac{x_{1}+x_{2}}{2}%
-R_{0}\right) ^{2}-\frac{\beta ^{2}}{2}\left( \left\vert
x_{1}-x_{2}\right\vert +r_{0}\right) ^{2} \\
&&+i\left( K_{0}\frac{x_{1}+x_{2}}{2}\mp k_{0}\left\vert
x_{1}-x_{2}\right\vert -k_{0}r_{0}+K_{0}R_{0}\right) .  \notag
\end{eqnarray}%
Furthermore, from the identity

\begin{eqnarray}
&&\alpha ^{2}\left( x_{1}+x_{2}+A\right) ^{2}+4\beta ^{2}\left(
x_{1}-x_{2}+B\right) ^{2}=\alpha ^{2}\left[ \left( x_{1}+A\right)
^{2}+\left( x_{2}+A\right) ^{2}-A^{2}\right] \\
&&+4\beta ^{2}\left[ \left( x_{1}+B\right) ^{2}+\left( x_{2}-B\right)
^{2}-B^{2}\right] +\left( \alpha ^{2}-4\beta ^{2}\right) x_{1}x_{2}  \notag
\end{eqnarray}%
we can see that the cross term $x_{1}x_{2}$\ vanishes if we take $\alpha
=2\beta $. The initial state can be written as a separable form

\begin{eqnarray}
\Psi \left( x_{1},x_{2},0\right) &=&\frac{\pi \beta ^{2}}{\sqrt{\Lambda }%
k_{0}}\left\{ \left( k_{0}+\gamma \right) \right. \left[ \varphi _{+}\left(
x_{1}\right) \varphi _{-}\left( x_{2}\right) u\left( x_{2}-x_{1}\right)
+\varphi _{+}\left( x_{2}\right) \varphi _{-}\left( x_{1}\right) u\left(
x_{1}-x_{2}\right) \right] \\
&&+\left( k_{0}-\gamma \right) \left. \left[ \varphi _{+}\left( x_{2}\right)
\varphi _{-}\left( x_{1}\right) u\left( x_{2}-x_{1}\right) +\varphi
_{+}\left( x_{1}\right) \varphi _{-}\left( x_{2}\right) u\left(
x_{1}-x_{2}\right) \right] \right\} ,  \notag
\end{eqnarray}%
where $u\left( x\right) $ is Heaviside step function and

\begin{equation}
\varphi _{\pm }\left( x\right) =\exp \left[ -\beta ^{2}\left( x\mp \frac{%
r_{0}}{2}\right) ^{2}+\frac{i}{2}\left( K_{0}\pm 2k_{0}\right) x\right] .
\label{Phi(+/-)}
\end{equation}%
In this case, $\Lambda $ can be obtained as
\begin{equation*}
\Lambda =\frac{4\pi ^{3}\beta ^{2}\left( k_{0}-\gamma \right) ^{2}}{k_{0}^{2}%
}.
\end{equation*}%
Without loss of generality we have set the initial center-of-mass coordinate
$R_{0}=0$ and dropped an overall phase $k_{0}r_{0}$. We note that functions $%
\varphi _{+}\left( x\right) $\ and $\varphi _{-}\left( x\right) $\ represent
Gaussian functions with centers at $r_{0}/2$ and $-r_{0}/2$, respectively.
Obviously, the probability contributions of $\varphi _{+}\left( x_{2}\right)
\varphi _{-}\left( x_{1}\right) u\left( x_{2}-x_{1}\right) $ and $\varphi
_{+}\left( x_{1}\right) \varphi _{-}\left( x_{2}\right) u\left(
x_{1}-x_{2}\right) $ are negligible under the condition $\beta r_{0}\gg 1$.
We then yield

\begin{equation}
\Psi \left( x_{1},x_{2},0\right) \approx \beta \sqrt{\frac{1}{2\pi }}\left[
\varphi _{+}\left( x_{1}\right) \varphi _{-}\left( x_{2}\right) +\varphi
_{+}\left( x_{2}\right) \varphi _{-}\left( x_{1}\right) \right] ,
\label{initial state_1}
\end{equation}%
which represents two-boson wavepacket state with the same width, group
velocity $K_{0}/2\pm k_{0}$, and location $\mp r_{0}/2$. Here the
renormalization factor has been readily calculated by Gaussian integral. So
far we have construct an expected initial state\ without using the
biothogonal basis set. The dynamics of two separated boson wavepackets can
be described\ by the time evolution as that in the conventional quantum
mechanics.

\subsection{Annihilating collision}

It is presumable that before the bosons start to overlap they move as free
particles with the center moving in their the group velocities $K_{0}/2\pm
k_{0}$ and the width spreading as function of\ time $\sqrt{\left( 4\beta
^{4}t^{2}+1\right) /\beta ^{2}}$. We concern the dynamic behavior after the
collision. To this end, we calculate the time evolution of the given initial
state, which can be expressed as

\begin{figure}[tbp]
\includegraphics[ bb=65 180 535 560, width=5.9 cm, clip]{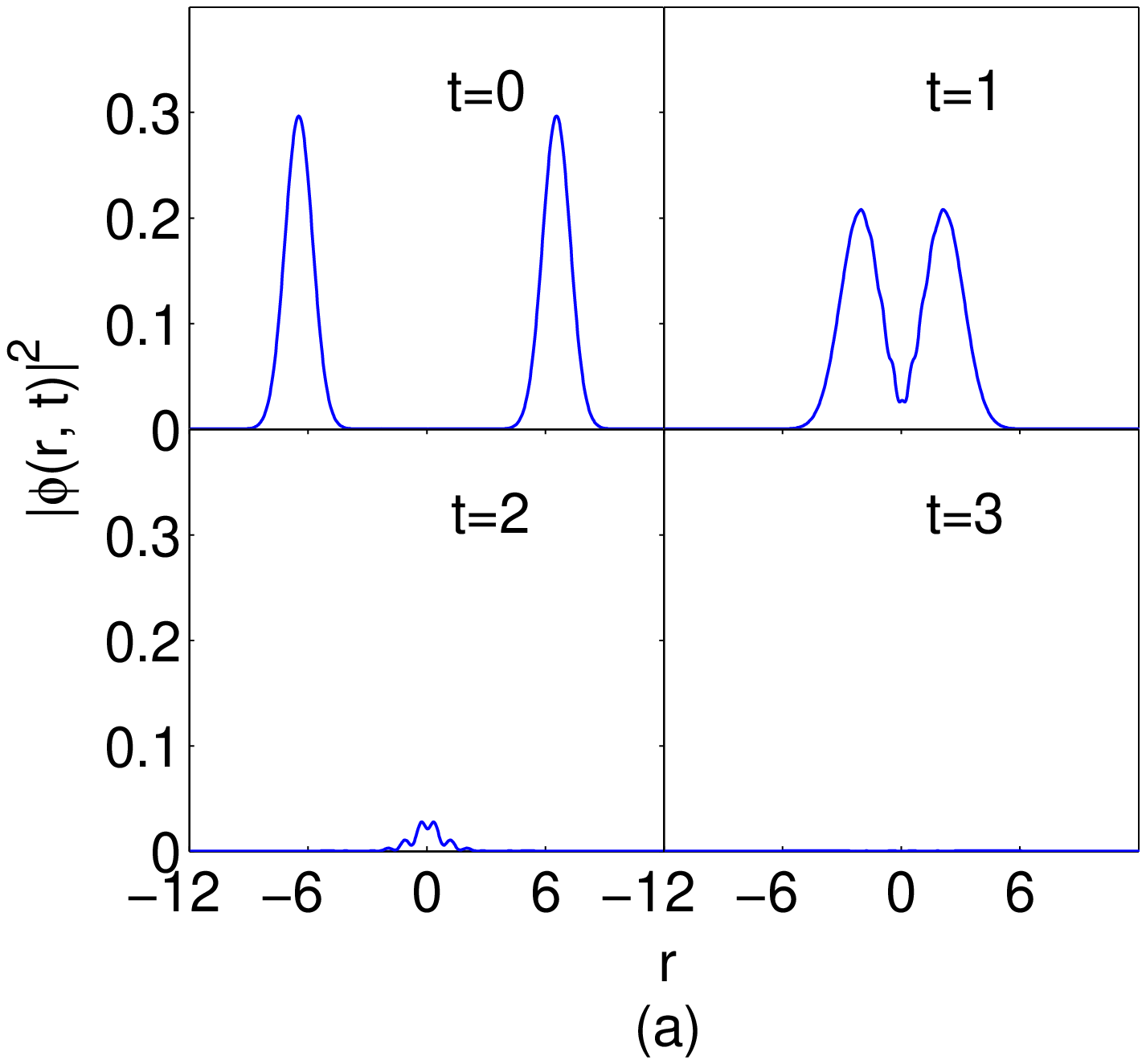} %
\includegraphics[ bb=65 180 535 560, width=5.9 cm, clip]{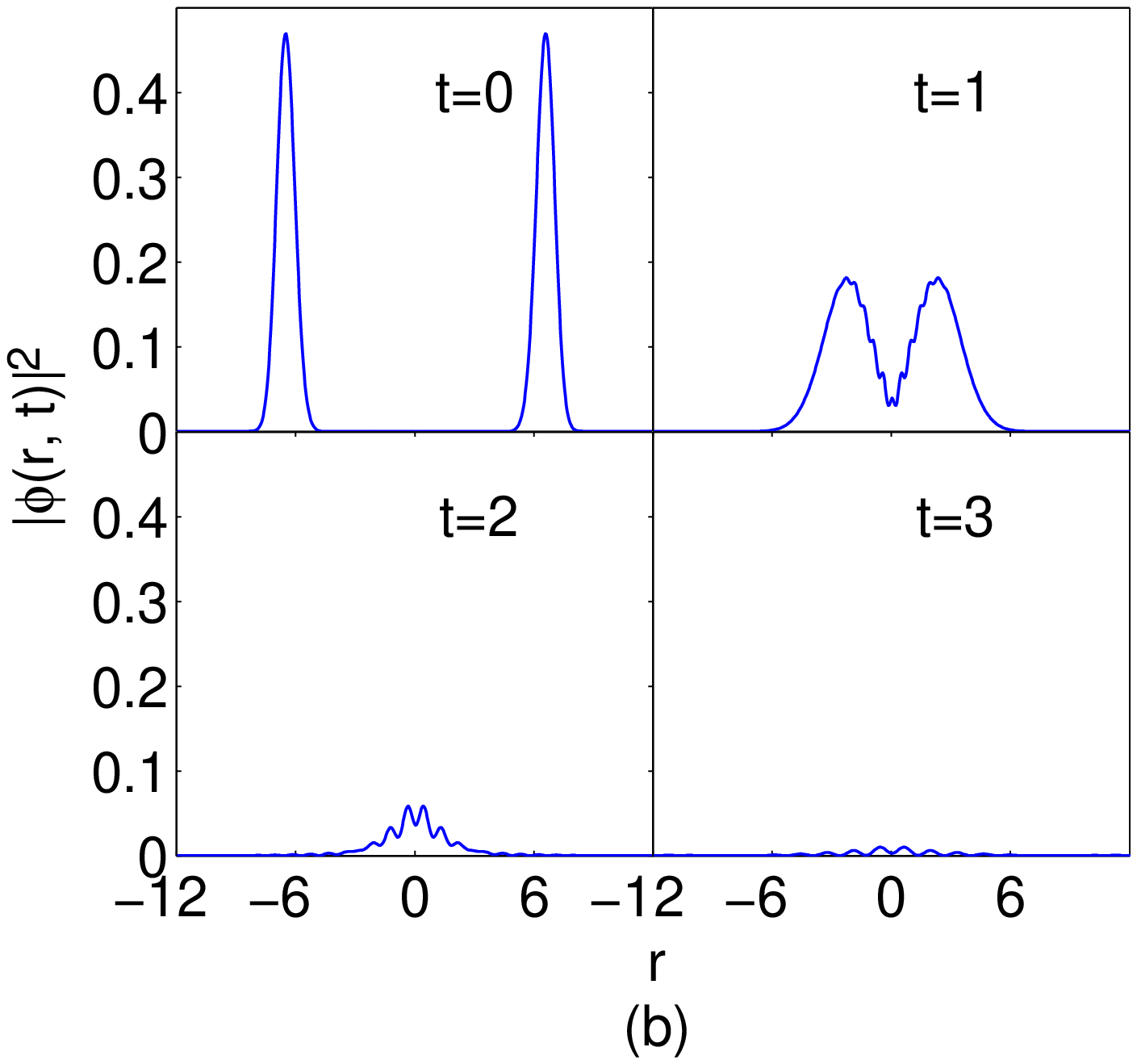} %
\includegraphics[ bb=65 180 535 560, width=5.9 cm, clip]{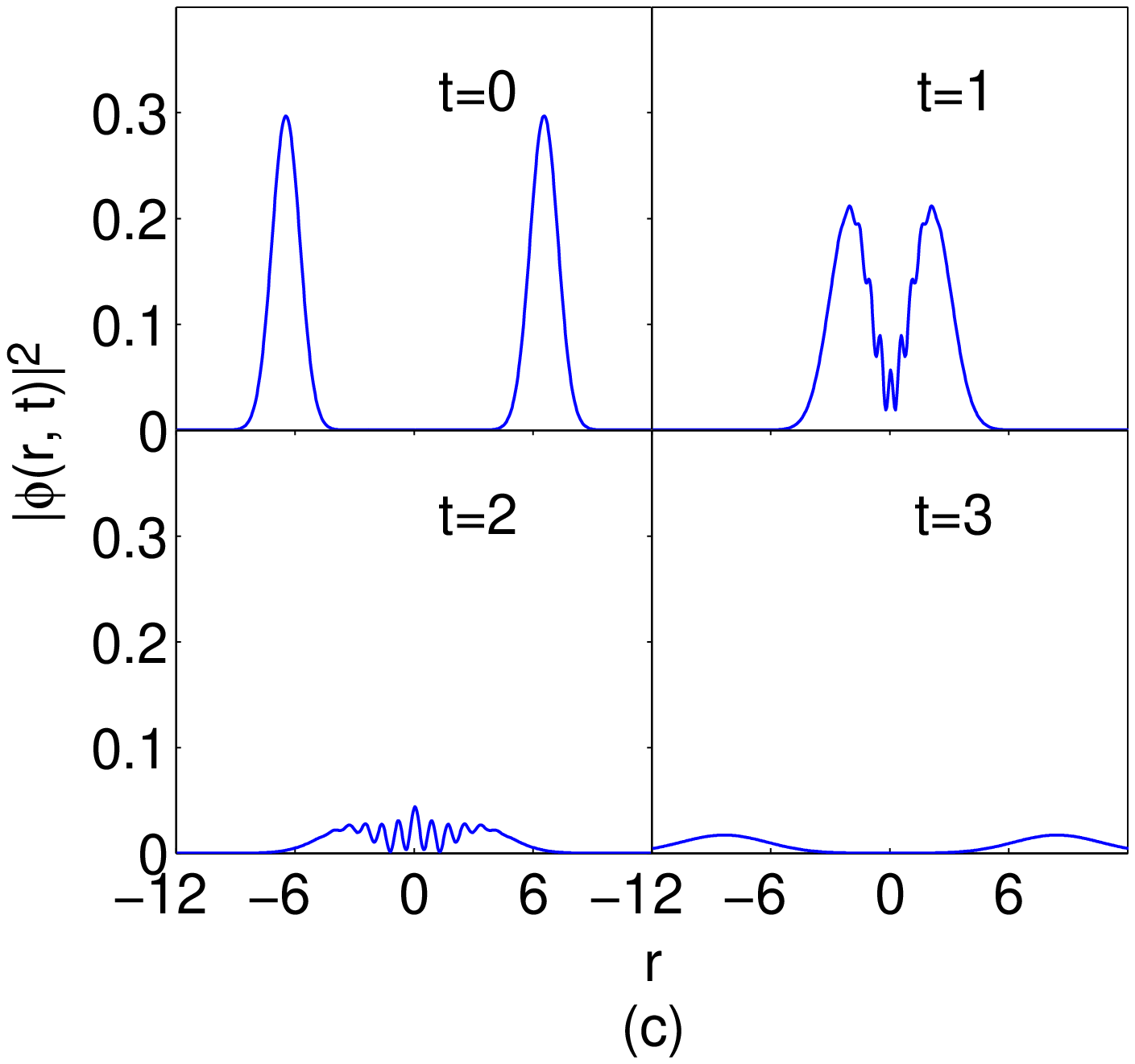}
\caption{(Color online) The profiles of $\left\vert \protect\varphi \left(
r,t\right) \right\vert ^{2}$ of the two-boson Gaussian wavepackets are
plotted for different values of $k_{0}$ and $\protect\alpha =2\protect\beta $%
: (a) $\protect\gamma =5.0,$ $k_{0}=5.0,\protect\alpha =2\protect\beta =2.0;$%
(b) $\protect\gamma =5.0,$ $k_{0}=5.0,\protect\alpha =2\protect\beta =3.0;$%
(c) $\protect\gamma =2.0,k_{0}=5.0,\protect\alpha =2\protect\beta =2.0.$ One
can see that the perfect pair annihilation in the case of (a) and imperfect
pair annihilation in the cases of (b) and (c) when the width of the initial
wavepackets becomes narrower, and the relative group velocity deviates from $%
\protect\gamma $, respectively. It shows that perfect pair annihilation can
be a signature of the singularity spectrum.}
\label{fig1}
\end{figure}

\begin{equation}
\Psi \left( x_{1},x_{2},t\right) =\frac{1}{\sqrt{\Lambda }}\int_{-\infty
}^{\infty }\int_{-\infty }^{\infty }G\left( K\right) g\left( k\right) \psi
\left( K,k,x_{1},x_{2}\right) e^{-i\left( K^{2}/4+k^{2}\right) t}\text{%
\textrm{d}}K\text{\textrm{d}}k.
\end{equation}%
By the similar procedure as above, we find that the evolved wave function
can always be written in the separated form%
\begin{equation}
\Psi \left( x_{1},x_{2},t\right) =\Phi \left( R,t\right) \phi \left(
r,t\right) ,
\end{equation}%
where%
\begin{equation}
\Phi \left( R,t\right) =\sqrt[4]{\frac{4\beta ^{2}}{\pi \left( 1+4\beta
^{2}t^{2}\right) }}\exp \left[ -\frac{2\beta ^{2}\left( R-K_{0}/2\right) ^{2}%
}{1+4\beta ^{2}t^{2}}+\frac{i\left( 16\beta
^{4}R^{2}+4RK_{0}-tK_{0}^{2}\right) }{4+16\beta ^{2}t^{2}}\right] ,
\end{equation}%
and\
\begin{equation}
\phi \left( r,t\right) =\frac{1}{\sqrt{\Omega }}\int_{-\infty }^{\infty
}\exp \left[ -\frac{1}{2\beta ^{2}}(k-k_{0})^{2}-ikr_{0}\right] \left[ \cos
\left( kr\right) -\frac{i\gamma }{k}\sin \left( k\left\vert r\right\vert
\right) \right] \exp \left( -ik^{2}t\right) \text{\textrm{d}}k.
\end{equation}%
where the normalization factor

\begin{equation}
\Omega =\frac{\pi ^{3/2}\beta \left( k_{0}-\gamma \right) ^{2}}{k_{0}^{2}}.
\end{equation}%
\ Straightforward algebra shows that%
\begin{equation}
\phi \left( r,t\right) =\left( k_{0}+\gamma \right) \Theta _{+}+\left(
k_{0}-\gamma \right) \Theta _{-}
\end{equation}%
where

\begin{eqnarray}
\Theta _{\pm } &=&\frac{\sqrt{\beta }\left\vert k_{0}-\gamma \right\vert
^{-1}}{\sqrt{2\pi ^{1/2}\left( 1+2i\beta ^{2}t\right) }}\exp \left\{ -\frac{%
\beta ^{2}\left[ \left\vert r\right\vert \pm \left( r_{0}-2k_{0}t\right) %
\right] ^{2}}{2\left( 4\beta ^{4}t^{2}+1\right) }+i\Delta _{\pm }\right\} ,
\\
\Delta _{\pm } &=&\frac{\beta ^{4}\left( \left\vert r\right\vert \pm
r_{0}\right) ^{2}t-2k_{0}^{2}t\mp 2k_{0}\left( \left\vert r\right\vert \pm
r_{0}\right) }{2\left( 4\beta ^{4}t^{2}+1\right) }.
\end{eqnarray}

In the case of $\beta ^{4}t^{2}\gg 1$, $k_{0}t\gg r_{0}$ the probability
distribution is

\begin{equation}
\left\vert \phi \left( r,t\right) \right\vert ^{2}\approx \frac{\pi \left(
k_{0}+\gamma \right) ^{2}}{4\Omega k_{0}^{2}t}\exp \left\{ -\frac{\left(
\left\vert r\right\vert +2k_{0}t\right) ^{2}}{4\beta ^{2}t^{2}}\right\} +%
\frac{\pi \left( k_{0}^{2}-\gamma ^{2}\right) e^{-k_{0}^{2}/\beta ^{2}}}{%
2\Omega k_{0}^{2}t}\exp \left( -\frac{r^{2}}{4\beta ^{2}t^{2}}\right) ,
\end{equation}%
which leads the total probability under the case $k_{0}/\beta \gg 1$
\begin{equation}
\int_{-\infty }^{\infty }\left\vert \phi \left( r,t\right) \right\vert ^{2}%
\text{\textrm{d}}r\approx \frac{\left( k_{0}+\gamma \right) ^{2}}{\left(
k_{0}-\gamma \right) ^{2}}.
\end{equation}

We can see that, after the collision the residual probability becomes a
constant and vanishes when $k_{0}=-\gamma $. It shows that when the relative
group velocity of two single-particle Gaussian wave packets with equal width
reaches the magnitude of the interaction strength, the dynamics exhibits
complete particle-pair annihilation.

In order to demonstrate such dynamic behavior and verify our approximate
result, the numerical method is employed to simulate the time evolution
process for several typical situations. The profiles of $\left\vert \varphi
\left( r,t\right) \right\vert ^{2}$ are plotted in Fig. 1. We would like to
point that the complete annihilation depends on the relative group velocity,
which is the consequence of singularity spectrum. This enhances the
probability of the pair annihilation for a cloud of bosons, which may\
provide an\ detection method of the spectral singularity in experiment.

\section{Second quantization representation}

\label{Second quantization representation}

In this section, we will investigate the two-particle collision process from
another point of view and give a more extended example. By employing the
second quantization representation, the initial state in Eq. (\ref{initial
state_1}) can be expressed as the form $a_{1}^{\dag }a_{2}^{\dag }\left\vert
0\right\rangle $, where $a_{i}^{\dag }$\ $\left( i=1,2\right) $\ is the
creation operator for a boson in single-particle state with the wavefunction
\begin{figure}[tbp]
\includegraphics[ bb=65 180 535 560, width=5.9 cm, clip]{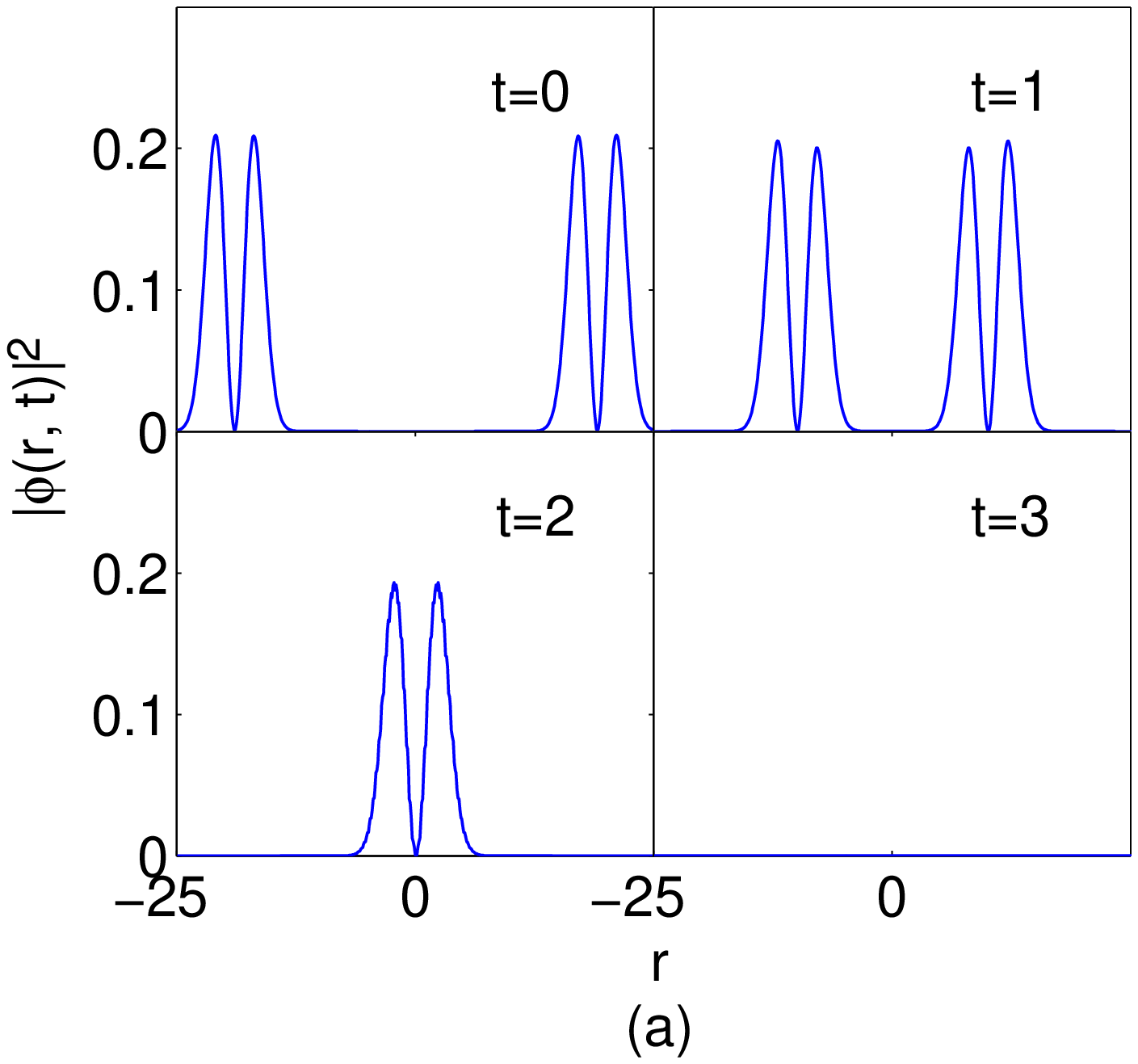} %
\includegraphics[ bb=65 180 535 560, width=5.9 cm, clip]{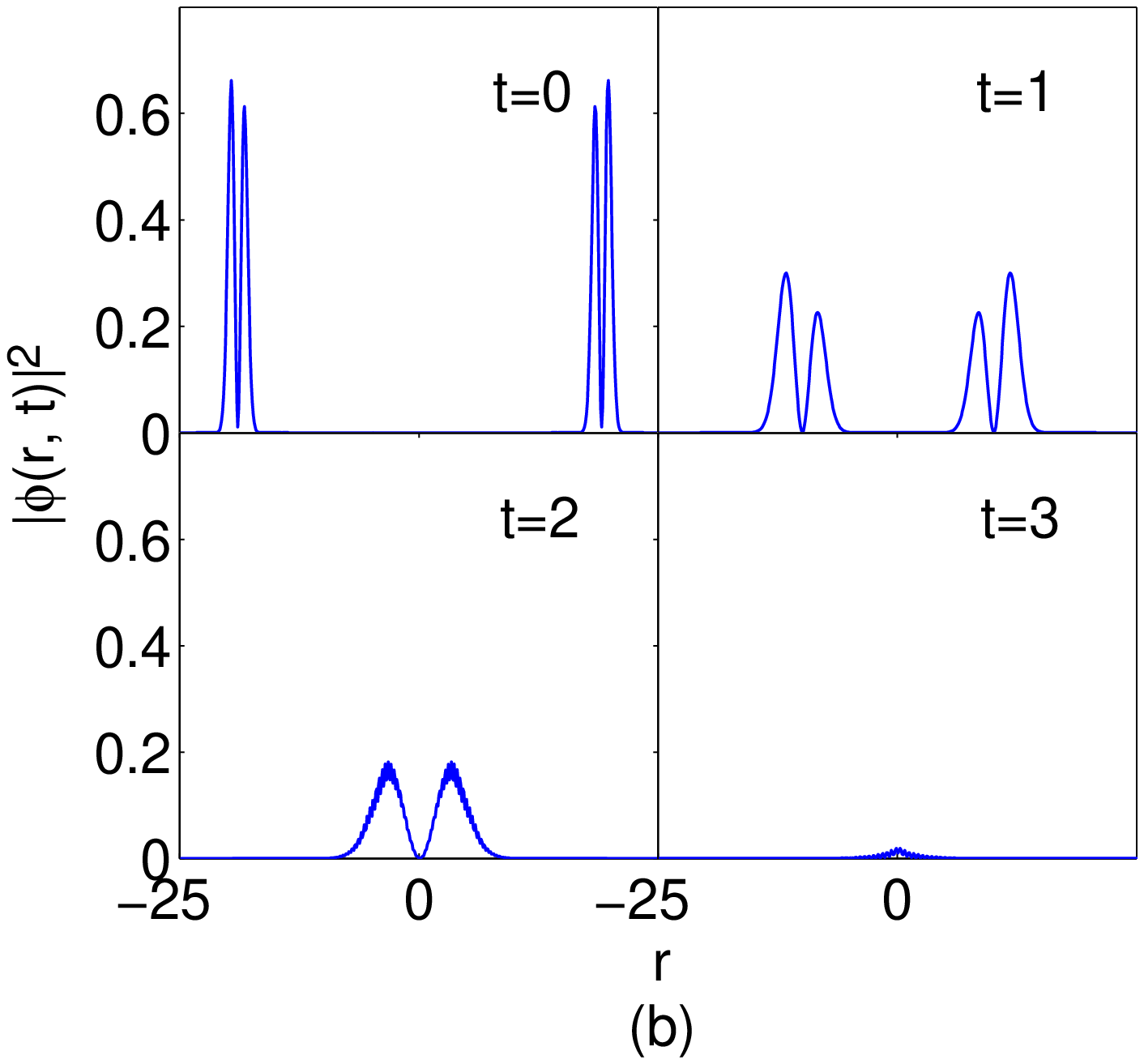} %
\includegraphics[ bb=65 180 535 560, width=5.9 cm, clip]{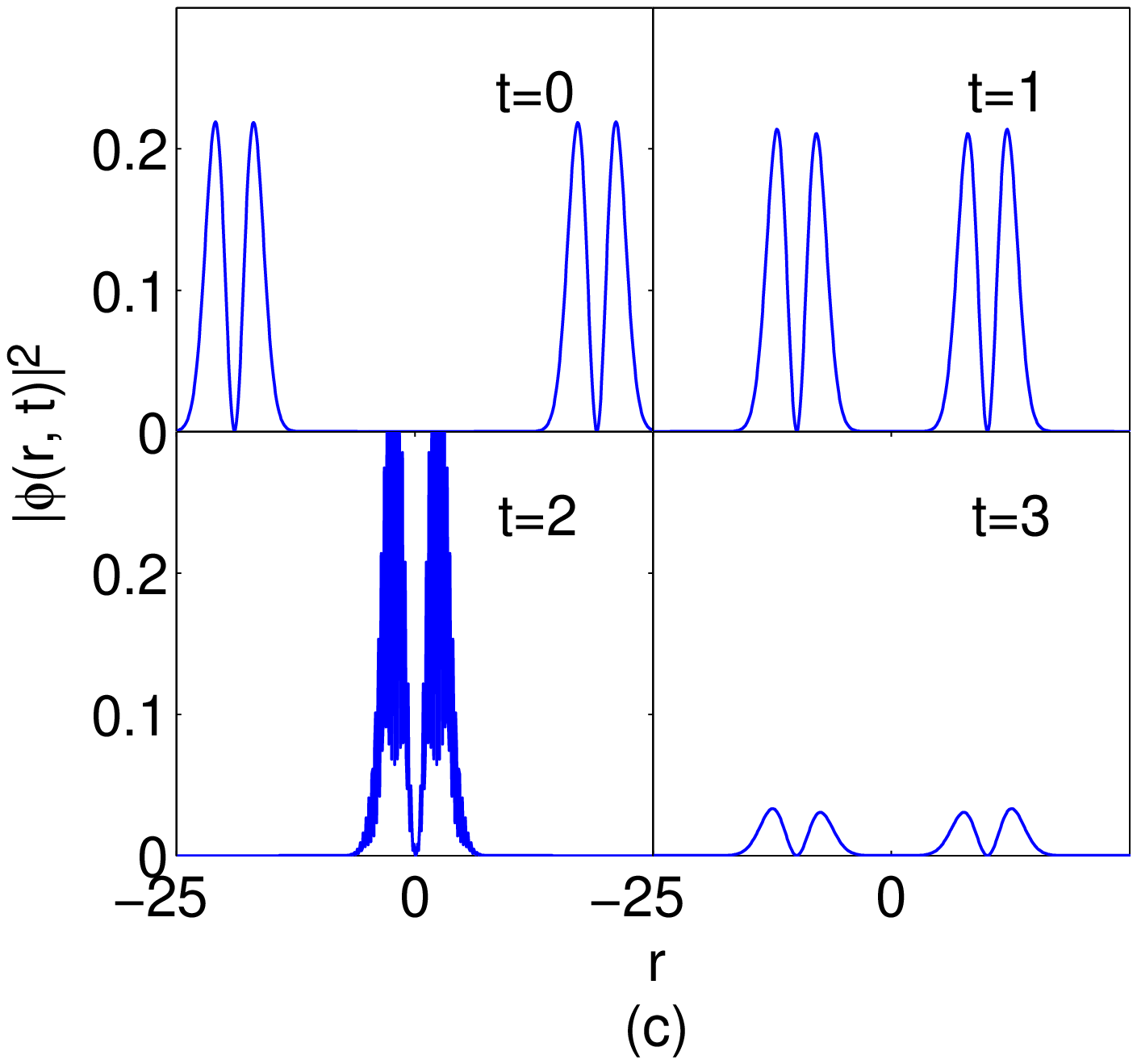}
\caption{(Color online) The profiles of $\left\vert \protect\varphi \left(
r,t\right) \right\vert ^{2}$ of a maximally two-mode entangled boson
pair are plotted for different values of $k_{0}$ and $\protect\alpha =2%
\protect\beta $: (a) $\protect\gamma =10.0,$ $k_{0}=10.0,\protect\alpha =2%
\protect\beta =1.0;$(b) $\protect\gamma =10.0,$ $k_{0}=10.0,\protect\alpha =2%
\protect\beta =3.0;$(c) $\protect\gamma =4.0,k_{0}=10.0,\protect\alpha =2%
\protect\beta =1.0.$ One can see that the perfect pair annihilation in the
case of (a) and imperfect pair annihilation in the cases of (b) and (c) when
the width of the initial wavepackets becomes narrower, and the relative
group velocity deviates from $\protect\gamma $, respectively.}
\label{fig2}
\end{figure}
\begin{equation}
\varphi _{+}\left( x\right) =\left\langle x\right\vert a_{1}^{\dag
}\left\vert 0\right\rangle \text{, }\varphi _{-}\left( x\right)
=\left\langle x\right\vert a_{2}^{\dag }\left\vert 0\right\rangle ,
\end{equation}%
and $\left\vert 0\right\rangle $\ denotes the vacuum state of the particle
operator. Similarly, if we consider a fermion pair, the initial state in Eq.
(\ref{initial state_1}) can be written as%
\begin{equation}
\frac{1}{\sqrt{2}}\left( c_{1,\uparrow }^{\dag }c_{2,\downarrow }^{\dag
}-c_{1,\uparrow }^{\dag }c_{2,\downarrow }^{\dag }\right) \left\vert
0\right\rangle ,  \label{fermion pair}
\end{equation}%
where $c_{i,\sigma }^{\dag }$ $\left( i=1,2;\sigma =\uparrow ,\downarrow
\right) $ is the creation operator for a fermion in single-particle state
with the wavefunction%
\begin{equation}
\varphi _{+}\left( x\right) \chi _{\sigma }=\left\langle x\right\vert
c_{1,\sigma }^{\dag }\left\vert 0\right\rangle \text{, }\varphi _{-}\left(
x\right) \chi _{\sigma }=\left\langle x\right\vert c_{2,\sigma }^{\dag
}\left\vert 0\right\rangle .
\end{equation}%
Here%
\begin{equation}
\chi _{\uparrow }=\binom{1}{0},\chi _{\downarrow }=\binom{0}{1},
\end{equation}%
are the spin part of wavefunction. We see that the initial state in Eq. (\ref%
{fermion pair}) is singlet pair with maximal entanglement. In contract,
state $\frac{1}{\sqrt{2}}\left( c_{1,\uparrow }^{\dag }c_{2\downarrow
}^{\dag }+c_{1,\uparrow }^{\dag }c_{2\downarrow }^{\dag }\right) \left\vert
0\right\rangle $\ should not lose any amplitude after collision.

On the other hand, we can extend our conclusion to other types of initial
state. For instance, we can construct the initial state with%
\begin{eqnarray}
G\left( K\right) &=&\exp \left[ -\frac{1}{2\alpha ^{2}}\left( K-K_{0}\right)
^{2}-iKR_{0}\right] , \\
g\left( k\right) &=&\left( k-k_{0}\right) \exp \left[ -\frac{1}{2\beta ^{2}}%
\left( k-k_{0}\right) ^{2}-ikr_{0}\right] ,  \notag
\end{eqnarray}%
which are also local states in $K$\ and $k$\ spaces, respectively.

In coordinate space, the above wavefunction has the from%
\begin{eqnarray}
\Psi \left( x_{1},x_{2},0\right)  &=&\frac{2i\pi \beta ^{4}}{\sqrt{\Xi }k_{0}%
}\left\{ \left( k_{0}+\gamma \right) \left[ \left( \varphi _{+}^{(1)}\left(
x_{1}\right) \varphi _{-}\left( x_{2}\right) -\varphi _{+}\left(
x_{1}\right) \varphi _{-}^{\left( 1\right) }\left( x_{2}\right) \right)
u\left( x_{2}-x_{1}\right) \right. \right.   \notag \\
&&+\left. \left( x_{1}\rightleftarrows x_{2}\right) \right]   \notag \\
&&-\left( k_{0}-\gamma \right) \left[ \left( \varphi _{-}^{\left( 1\right)
}\left( x_{1}\right) \varphi _{+}\left( x_{2}\right) -\varphi _{+}^{\left(
1\right) }\left( x_{2}\right) \varphi _{-}\left( x_{1}\right) \right)
u\left( x_{2}-x_{1}\right) \right.   \notag \\
&&+\left. \left. \left( x_{1}\rightleftarrows x_{2}\right) \right] \right\} ,
\end{eqnarray}%
which can be reduced to%
\begin{equation}
\Psi \left( x_{1},x_{2},0\right) \approx \frac{2i\pi \beta ^{4}\left(
k_{0}-\gamma \right) }{\sqrt{\Xi }k_{0}}\left[ \left( \varphi
_{+}^{(1)}\left( x_{1}\right) \varphi _{-}\left( x_{2}\right) -\varphi
_{+}\left( x_{1}\right) \varphi _{-}^{(1)}\left( x_{2}\right) \right)
+\left( x_{1}\rightleftarrows x_{2}\right) \right] ,  \label{initial state_2}
\end{equation}%
under the approximation $\beta r_{0}\gg 1$. Here $\Xi $ is the normalized
constant, and $\varphi _{\pm }^{(1)}\left( x\right) =\left( x\mp \frac{r_{0}%
}{2}\right) \varphi _{\pm }\left( x\right) $.

By the same procedure, at time $t$ the evolved wavefunction is%
\begin{equation}
\phi \left( r,t\right) =-\left( k_{0}+\gamma \right) \Theta _{+}+\left(
k_{0}-\gamma \right) \Theta _{-},
\end{equation}%
with%
\begin{eqnarray}
\Theta _{\pm } &=&\sqrt{\frac{\pi }{2\Omega ^{\prime }}}\frac{i\beta ^{3}%
\left[ \left\vert r\right\vert \pm \left( r_{0}-2k_{0}t\right) \right] }{%
k_{0}\left( 1+2i\beta ^{2}t\right) ^{3/2}}\exp \left\{ -\frac{\beta ^{2}%
\left[ \left\vert r\right\vert \pm \left( r_{0}-2k_{0}t\right) \right] ^{2}}{%
2\left( 4\beta ^{4}t^{2}+1\right) }+i\Delta _{\pm }\right\} , \\
\Delta _{\pm } &=&\frac{\beta ^{4}\left( \left\vert r\right\vert \pm
r_{0}\right) ^{2}t-2k_{0}^{2}t\mp 2k_{0}\left( \left\vert r\right\vert \pm
r_{0}\right) }{2\left( 4\beta ^{4}t^{2}+1\right) }.
\end{eqnarray}%
where the normalization factor%
\begin{equation*}
\Omega ^{\prime }=\frac{\pi ^{3/2}\beta ^{3}\left( k_{0}-\gamma \right) ^{2}%
}{4k_{0}^{2}}.
\end{equation*}%
In the case of $\beta ^{4}t^{2}\gg 1$, $k_{0}t\gg r_{0}$ the probability
distribution is

\begin{eqnarray}
\left\vert \phi \left( r,t\right) \right\vert ^{2} &\approx &\frac{\pi
\left( k_{0}+\gamma \right) ^{2}}{16\Omega ^{\prime }k_{0}^{2}t^{3}}\left(
\left\vert r\right\vert +2k_{0}t\right) ^{2}\exp \left\{ -\frac{\left(
\left\vert r\right\vert +2k_{0}t\right) ^{2}}{4\beta ^{2}t^{2}}\right\} \\
&&-\frac{\pi \left( k_{0}^{2}-\gamma ^{2}\right) e^{-k_{0}^{2}/\beta ^{2}}}{%
8\Omega ^{\prime }k_{0}^{2}t^{3}}\left( r^{2}-4k_{0}^{2}t^{2}\right) \exp
\left( -\frac{r^{2}}{4\beta ^{2}t^{2}}\right) ,  \notag
\end{eqnarray}%
which leads the total probability%
\begin{equation}
\int_{-\infty }^{\infty }\left\vert \phi \left( r,t\right) \right\vert ^{2}%
\text{\textrm{d}}r\approx \frac{\left( k_{0}+\gamma \right) ^{2}}{\left(
k_{0}-\gamma \right) ^{2}}.
\end{equation}%
The profiles of $\left\vert \varphi \left( r,t\right) \right\vert ^{2}$ are
plotted in Fig. 2. We can see that the same behavior occurs in the present
situation.

In order to clarify the physical picture, we still employ the second
quantization representation by introducing another type of boson creation
operator $b_{i}^{\dag }$\ $\left( i=1,2\right) $\ with%
\begin{eqnarray*}
\varphi _{-}^{(0)}\left( x\right) &=&\left\langle x\right\vert b_{1}^{\dag
}\left\vert 0\right\rangle , \\
\varphi _{-}^{(0)}\left( x\right) &=&\left\langle x\right\vert b_{2}^{\dag
}\left\vert 0\right\rangle .
\end{eqnarray*}%
Then the initial state in Eq. (\ref{initial state_2}) can be expressed as%
\begin{equation}
\frac{1}{\sqrt{2}}\left( a_{1}^{\dag }b_{2}^{\dag }+a_{2}^{\dag }b_{1}^{\dag
}\right) \left\vert 0\right\rangle ,
\end{equation}%
which is maximally two-mode entangled state.

\bigskip

\section{Summary and discussion}

\label{Summary}

In summary we identified a connection between spectral singularities and
dynamical behavior for interacting many-particle system. We explored the
collision process of two bosons as well as fermions in one-dimensional
system with imaginary delta interaction strength based on the exact
solution. We have showed that there is a singularity spectrum which leads to
complete particle-pair annihilation when the relative group velocity is
resonant to the magnitude of interaction strength. The result for this
simple model implies that the complete particle-pair annihilation can only
occur for two distinguishable bosons, maximally two-mode entangled boson
pair and singlet fermions, which may predict the existence of its
counterpart in the theory of particle physics.

\acknowledgments We acknowledge the support of the National Basic Research
Program (973 Program) of China under Grant No.2012CB921900 and CNSF (Grant
No. 11374163).


\begin{thebibliography}{99}
\bibitem{Gamow} G. Gamow, Z. Phys. A \textbf{51}, 204 (1928).

\bibitem{Dattoli} G. Dattoli, A. Torre, and R. Mignani, Phys. Rev. A \textbf{%
42}, 1467 (1990).

\bibitem{Okolowicz} J. Okolowicz, M. Ploszajczak, and I. Rotter, Phys. Rep.
\textbf{374}, 271 (2003).

\bibitem{Moiseyev} N. Moiseyev, Phys. Rep. \textbf{302}, 212 (1998).

\bibitem{Muga} J. G. Muga, J. P. Palao, B. Navarro, and I. L. Egusquiza
Phys. Rep. \textbf{395}, 357 (2004).

\bibitem{Bender} C. M. Bender and S. Boettcher, Phys. Rev. Lett. \textbf{80}%
, 5243 (1998).

\bibitem{Ann} F. G. Scholtz, H. B. Geyer and F. J. W. Hahne, Ann. Phys. (NY)
\textbf{213}, 74 (1992).

\bibitem{JMP1} C. M. Bender, S. Boettcher and P. N. Meisinger, J. Math.
Phys. \textbf{40}, 2201 (1999).

\bibitem{PRL1} C. M. Bender, D. C. Brody, and H. F. Jones, Phys. Rev. Lett.
\textbf{89}, 270401 (2002).

\bibitem{JPA1} A. de Souza Dutra, M. B. Hott and V. G. C. S. dos Santos, J.
Phys. A \textbf{34}, L 391 (2001).

\bibitem{JPA2} P. Dorey, C. Dunning and R. Tateo, J. Phys. A \textbf{34},
5679 (2001).

\bibitem{JMP2} A. Mostafazadeh, J. Math. Phys. \textbf{43}, 205 (2002).

\bibitem{JMP3} A. Mostafazadeh, J. Math. Phys. \textbf{43}, 2814 (2002).

\bibitem{JMP4} A. Mostafazadeh, J. Math. Phys. \textbf{43}, 3944 (2002).

\bibitem{JPA3} A. Mostafazadeh and A. Batal, J. Phys. A \textbf{36}, 7081
(2003).

\bibitem{JPA4} A. Mostafazadeh and A. Batal, J. Phys. A \textbf{37}, 11645
(2004).

\bibitem{JPA5} H. F. Jones, J. Phys. A \textbf{38}, 1741 (2005).

\bibitem{PRA1} A. Mostafazadeh, Phys. Rev. A \textbf{80}, 032711 (2009).

\bibitem{PRA3} A. Mostafazadeh, Phys. Rev. A \textbf{84}, 023809 (2011).

\bibitem{Ali3} A. Mostafazadeh, Phys. Rev. Lett. \textbf{110}, 260402 (2013).

\bibitem{PRA4} A. Mostafazadeh and M. Sarisaman, Phys. Rev. A \textbf{87},
063834 (2013).

\bibitem{PRA6} A. Mostafazadeh and M. Sarisaman, Phys. Rev. A \textbf{88},
033810 (2013).

\bibitem{PRB1} S. Longhi, Phys. Rev. B \textbf{80}, 165125 (2009).

\bibitem{JMP5} A. A. Andrianov, F. Cannata and A. V. Sokolov, J. Math. Phys.
\textbf{51}, 052104 (2010).

\bibitem{PRD1} F. Correa and M. S. Plyushchay, Phys. Rev. D \textbf{86},
085028 (2012).

\bibitem{PRA5} L. Chaos-Cador and G. Garca-Caldern Phys. Rev. A \textbf{87},
042114 (2013).

\bibitem{PRL3} A. Mostafazadeh, Phys. Rev. Lett. \textbf{102}, 220402 (2009).

\bibitem{prd2} H. F. Jones, Phys. Rev. D \textbf{76}, 125003 (2007).

\bibitem{prd7} H. F. Jones, Phys. Rev. D \textbf{78}, 065032 (2008).

\bibitem{prd6} M. Znojil, Phys. Rev. D \textbf{78}, 025026 (2008).

\bibitem{prd3} M. Znojil, Phys. Rev. D \textbf{80}, 045009 (2009).

\bibitem{prd4} M. Znojil, Phys. Rev. D \textbf{80}, 045022 (2009).

\bibitem{prd5} M. Znojil, Phys. Rev. D \textbf{80}, 105004 (2009).

\bibitem{prd8} C. M. Bender and P. D. Mannheim, Phys. Rev. D \textbf{78},
025022 (2008).

\bibitem{PRA2} S. Longhi, Phys. Rev. A \textbf{81}, 022102 (2010).

\bibitem{JPA6} A. Ghatak, J. A. Nathan, B. P. Mandal and Z. Ahmed, J. Phys.
A: Math. Theor. \textbf{45}, 465305 (2012).

\bibitem{PRA13} A. Mostafazadeh, Phys. Rev. A \textbf{87}, 063838 (2013).

\bibitem{PRA14} X. Z. Zhang, L. Jin, and Z. Song, Phys. Rev. A \textbf{87},
042118 (2013).

\bibitem{Ali1} A. Mostafazadeh, J. Phys. A: Math. Gen. \textbf{39},
13495--13506 (2006).

\bibitem{Samsonov} Samsonov B F, J. Phys. A: Math. Gen. \textbf{38}, L 571
(2005).
\end{thebibliography}
\end{document}